\documentclass[prl,twocolumn,showpacs,superscriptaddress]{revtex4}
\usepackage{graphicx}

\begin{document}

\title {Geometry-dependent scattering through quantum billiards:\\
  Experiment and theory}

\author{T. Blomquist}
\affiliation{Department of Physics (IFM),
Link\"{o}ping University, S--581\,83 Link\"{o}ping, Sweden}
\author{H. Schanze}
\affiliation{Fachbereich Physik, Philipps-Universit\"at
Marburg D-35032 Marburg, Germany}
\author{I. V. Zozoulenko}
\affiliation{
Department of Science and Technology (ITN), Link\"{o}ping University,
S--601\,74
Norrk\"oping, Sweden}
\affiliation{Department of Physics (IFM),
Link\"{o}ping University, S--581\,83 Link\"{o}ping, Sweden}
\author{H.-J. St\"ockmann}
\affiliation{Fachbereich Physik, Philipps-Universit\"at
Marburg D-35032 Marburg, Germany}

\date{\today}

\begin{abstract}
We present experimental studies of the geometry-specific quantum
scattering in microwave billiards of a given shape. We perform
full quantum mechanical  scattering
calculations and find an excellent agreement with the experimental
results. We also carry out the semiclassical  calculations where the
conductance is 
given as a sum of all classical trajectories between the leads,
each of them carrying the quantum-mechanical phase. We
unambiguously demonstrate that the characteristic frequencies of
the oscillations in the transmission and
reflection \emph{amplitudes} are related to the \emph{length
distribution} of the classical trajectories between the leads,
whereas the frequencies of the \emph{probabilities} can be
understood in terms of the \emph{length difference distribution}
in the pairs of classical trajectories. We also discuss the effect
of non-classical ``ghost'' trajectories that include classically
forbidden reflection off the lead mouths.

\end{abstract}

\maketitle

The low-temperature conductance of  nanoscaled semiconductor quantum
dots (often called quantum  billiards) is dominated
by  quantum mechanical interference of electron waves giving
rise to reproducible conductance oscillations
\cite{Marcus,Chang,Persson,Bird,Ensslin,comment,Christ,Bogg,andy_sq,Z_PRL}.
Theoretical and experimental studies of the conductance oscillations
have been concentrated on both statistical and geometry-specific
features
\cite{Marcus,Chang,Persson,Bird,Ensslin,comment,Christ,Bogg,andy_sq,Z_PRL,%
Takagaki,RMT,semicl,Delos,Burg,TB1,TB2}.
The analysis of the statistical aspects of the conductance
was commonly based on the random matrix theory or similar stochastic
methods \cite{RMT}.  In order to provide
the interpretation of the geometry-specific features in the oscillations
in a billiard of a given shape, different and sometimes conflicting
approaches have been used
\cite{Marcus,Chang,Persson,Bird,Ensslin,comment,Christ,Bogg,andy_sq,Z_PRL,%
Takagaki,Delos,Burg,TB1,TB2}.
Very often the interpretation of the
conductance was not directly based on the transport calculations such that the
 explanation of the
characteristic peaks in the conductance spectrum  had rather speculative
character.
In contrast, the semiclassical approach
\cite{semicl,Delos,Burg,TB1,TB2,Gutzwiller,stock_book} represents one of the
most powerful tool for studying the geometry-specific scattering
as it allows one to perform transport calculations for structures of
arbitrary geometry
and, at the same time, it can provide an intuitive interpretation of
the conductance in terms of classical trajectories connecting the
leads, each of them  carrying the
quantum mechanical phase.

 In this paper we, for the first time, present experimental studies of
 the geometry-specific
quantum scattering in microwave billiards of a given shape combined
with the exact quantum mechanical as well as semi-classical analysis. The
physics and  modelling  of microwave cavities is conceptually similar
to that one of the semiconductor quantum dots  because of the
similarity between the Schr\"odinger and Helmholtz equations
\cite{stock_book,bird_paper}. At the
same time, the microwave cavities provide the unique opportunity to
control the precise shape of the billiard. This is not
possible for the semiconductor quantum dots where the actual shape
of the potential always remains unknown \cite{potential}.  Secondly,
 for the microwave
billiards one can routinely access the phase information (i.e.
measuring the transmission and reflection
\emph{amplitudes}) whereas for the case of quantum dot one typically
measures the transmission \emph{probabilities} only. Thirdly, the effects
of inelastic scattering is of much lesser importance for the
microwave cavities than for the quantum dots where the phase
breaking processes can reduce the phase coherence length
significantly. Note that a billiard system of the same shape as
the one studied here (realized as a semiconductor quantum dot),
was investigated in \cite{Ensslin,TB1}. It has been demonstrated
\cite{TB1} that the strong inelastic scattering has lead to the
suppression of major characteristic peaks in the transmission
spectrum as well as to the strong reduction of the amplitude of
the oscillations.

Our semiclassical (SC) analysis of the experimental
spectrum of the microwave billiard unambiguously demonstrate that
the characteristic frequencies of the oscillations in the
transmission and reflection \emph{amplitudes} are related
to the \emph{length distribution} of the classical trajectories
between the leads, whereas the frequencies of the
\emph{probabilities} can be understood in terms of the
\emph{length difference distribution} in the pairs of classical
trajectories. This, to the best of our knowledge, provides the
first {\em unambiguous} identification of the specific frequencies
of the oscillations experimentally observed in a billiard of a
given shape.

The dynamics of an electron in a two-dimensional quantum dot is
governed by the Schr\"odinger equation
\begin{equation}
\left(\frac{\hbar^2}{2m}\nabla^2+E\right)\psi (x,y)=0,
\end{equation}
with the wave function vanishing on
the boundary \textbf{$\psi=0$}, where $E$ is the electron Fermi
energy and a potential inside the billiard is assumed to be zero.
This equation has the same form as the Helmholtz equation
governing the dynamics of the lowest TM-mode in a microwave
billiards \cite{stock_book} (provided the frequency $\nu<c/2d$,
where $d$ is the heigt of the billiard).

%-------------------------- Fig 0--------------------------
\begin{figure}
\includegraphics[width=0.3\textwidth]{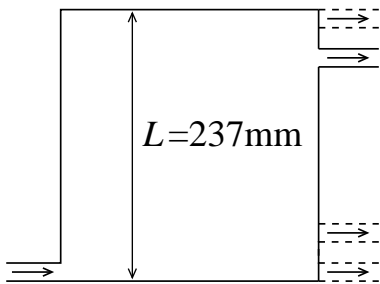}
\caption{\label{fig0}Sketch of the resonator used in the
experiment (in scale). Measurements have been taken for four
different positions of the outgoing wave guide as indicated in the
figure.
}
\end{figure}
%----------------------------------------------------------

 In the absence of a magnetic
field the transmission amplitude $t_{nm}$ is given by the
projection of the retarded Green function $G=(H-E)^{-1}$ onto the
transverse wave functions $\phi _n(y)$ in the incoming and
outgoing leads \cite{Datta}
%********************Eq. (2)******************************************
\begin{eqnarray}\label{t}
t_{mn}(k)&=&
-i\hbar \sqrt{v_nv_m}\\ \nonumber
&\times&
\int dy_1\int dy_2 \phi_n^{*} (y_1) \phi_m
(y_2) G(y_1,y_2,k),
\end{eqnarray}
%********************Eq. (2)******************************************
where $v_{n}$ is the longitudinal velocity for the mode $n$ and $k$ is the
wave vector.   The total
transmission probability $T=\sum _{mn} T_{mn}$ is the sum over transmission
probabilities from all the propagating modes $m$ in one lead to the
modes $n$ in
the other; $T_{mn}$ being the square
modulus of the transmission amplitude, $T_{mn}= |t_{mn}|^2$.

The quantum mechanical computations have been made using a recursive
Green's function technique based on the Dyson equation \cite{recgreen}.
In semi-classical computations, the quantum mechanical Green's
function is replaced by its
semiclassical approximation \cite{Gutzwiller,stock_book}. The
semiclassical transmission
amplitude  can
be represented in the form \cite{Delos,Burg,TB1,TB2}
\begin{equation}\label{amp}
t^{SC}_{mn}(k)=\sum_s A_{mn}^{s} e^{ik l_s},
\end{equation}
where $s$ denotes classical trajectories  of the length $l_s$ between the
two leads, $A_{mn}^s$
is an amplitude factor which depends on the density of trajectories,
mode number, entrance and exit angles, etc. The details of the
semiclassical calculations are given in \cite{TB2}.

Because of the rapidly varying phase factor in Eq. (\ref{amp}), the length
spectrum of the transmission amplitude (defined as a Fourier Transform
$\int\! d\ell\; t^{SC}_{mn}(k)e^{-k\ell}$),
is obviously peaked at  the length $\ell=l_s$ of the trajectories
between the leads. This behavior of the length spectrum is well understood
and numerically verified for
a number of different model billiards with leads \cite{Delos,Burg,TB1}.

Using Eq. (\ref{amp}), the transmission probability can be
written in the form
\begin{equation}\label{probab}
T^{SC}_{mn}=|t^{SC}_{mn}|^2 =\sum_s |A_{mn}^{s}|^2 +
\sum_{s,s'}A_{mn}^{s}\, {A_{mn}^{s'}}^{*}   e^{ik(l_s-l_{s'})}.
\end{equation}
The length spectrum of the transmission probability is
obviously peaked at the length difference
$\ell=l_s-l_{s'}$ in the pairs of trajectories between the leads.
Thus, identification of the characteristic frequencies in the
probabilities reduces to the analysis of the path
difference distribution in a billiard with a given lead
geometry\cite{Delos,TB2}.

Figure \ref{fig0} shows a sketch of the microwave
resonator used in the experiments. The microwaves enter the
resonator through a wave guide at a fixed position on one side,
and leaves the resonator through another waveguide on the opposite
side whose position could be varied. The heigth of the resonator
was $d=7.8$ mm, i.e. the billiard is quasi-two-dimensional for
$\nu<19$ GHz. Transmision spectra, including modulus and phase,
were taken in the frequency range $10$ to $18$ GHz for four
different positions of the outgoing wave guide. In the whole
frequency range there is only one propagating mode in the wave
guide. More details on the experimental technique may be found in
\cite{Kuhl}.
%-------------------------- Fig 1--------------------------
\begin{figure}
\includegraphics[width=0.4\textwidth]{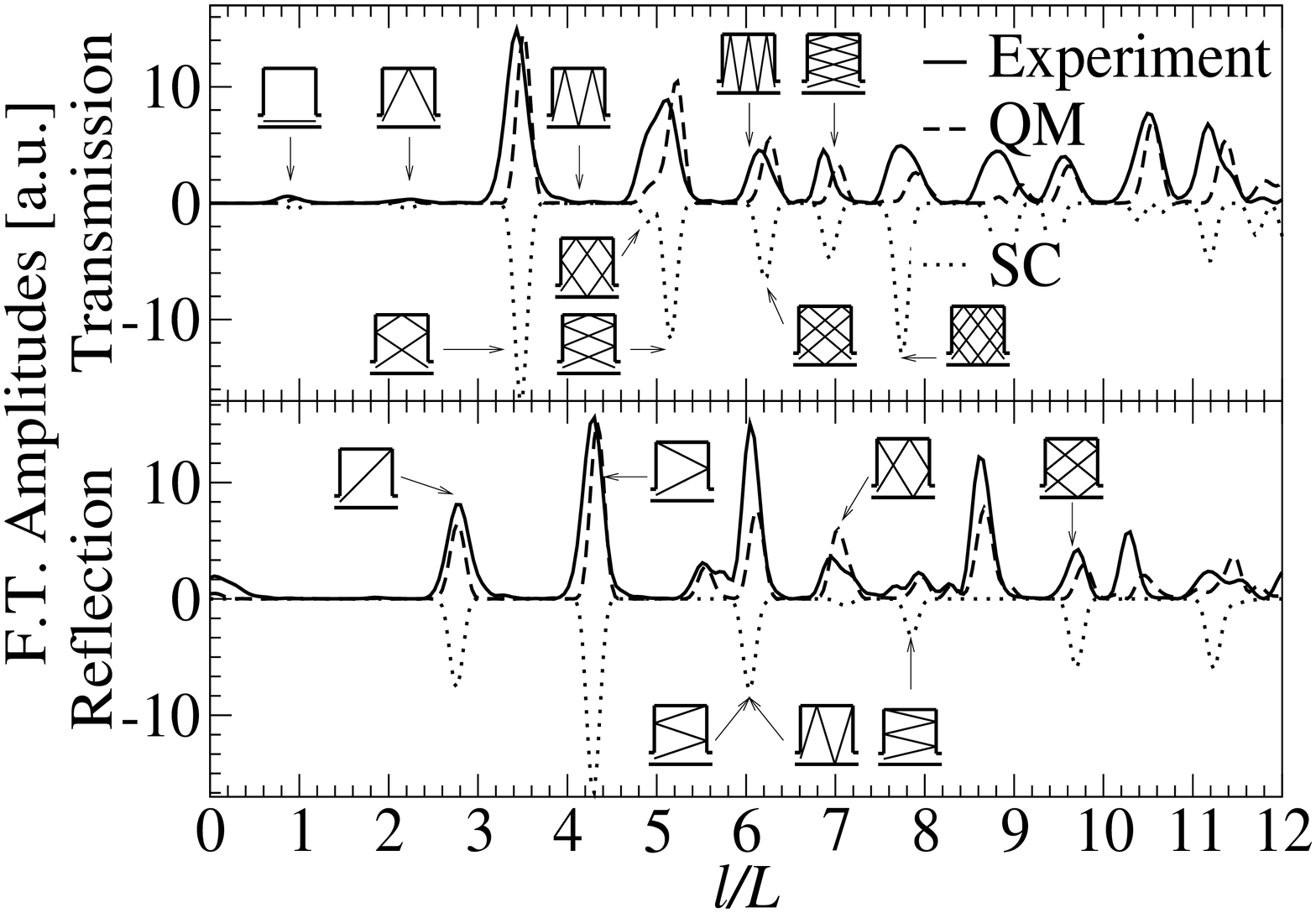}
\caption{\label{fig1} Fourier transform of the
experimental and calculated quantum mechanical (QM)
 transmission and  reflection \textit{amplitudes} for the square
 billiard with  opposite leads. The lower curve shows corresponding
 semiclassical (SC)
  results, plotted with a negative sign for the sake of clearness. The
  characteristic 
  peaks are identified in terms of classical transmitted and reflected
  trajectories. Peaks in
  the reflection amplitude at $l\approx 5.5L, \ 8.8L$ are due to
  ``ghost'' trajectories  that include classically forbidden reflection
off the lead mouths. The range of the frequency variation
corresponds to one propagating mode in the leads. The
insets show the schematic geometry of the experimental microwave
billiard. }
\end{figure}
%----------------------------------------------------------

Figure \ref{fig1} shows the experimental and calculated data for
the transmission and reflection \emph{amplitudes}. The agreement
between the experimental results and exact quantum
mechanical (QM) calculations is very good. The SC transport
calculations allows us to identify the characteristic peaks in the
spectrum in terms of classical trajectories connecting the
billiard leads. Indeed, each peak in the SC spectrum represents a
contribution from a particular classical trajectory, as
illustrated in the insets. However, because
  of the approximate nature of the semiclassical approximation,  the
  heights of the SC and QM peaks do not fully agree with each other.

  Furthermore, the experimental data as well as the  QM
  calculations show the presence of the peaks which are absent in the
  SC calculations (for example, the peaks at $l\approx 5.5, 9,
  10.5$ in the reflection amplitude). These are so-called
``ghost'' trajectories that include classically forbidden reflection
off the lead mouths \cite{Delos}.  For example, the peak in
  the reflection amplitude at $l\approx 9$ is caused by the trajectory
  with the length $l=4.5$ which, after one revolution
  in the billiard, is  reflected
  back at the exit by the lead mouth such that it makes one more
  revolution and its total length becomes
  $L=4.5\times 2\approx 9$. Such non-classical trajectories are not included
in the standard semiclassical approximation.

The ``ghost''
trajectories are more important for the reflection than for the
  transmission. This is due to the fact that each
  ``ghost'' trajectory manifesting itself in the reflection
  bounces  off the lead mouth
  only once, whereas each ``ghost''
  trajectory contributing to the transmission has to bounce off the lead mouth
  twice. As a result, the amplitude of such the trajectory with two
  non-classical bounces off the lead mouth is obviously lower than the
  one with only one bounce.

%-------------------------- Fig 2--------------------------
\begin{figure}
\includegraphics[width=0.4\textwidth]{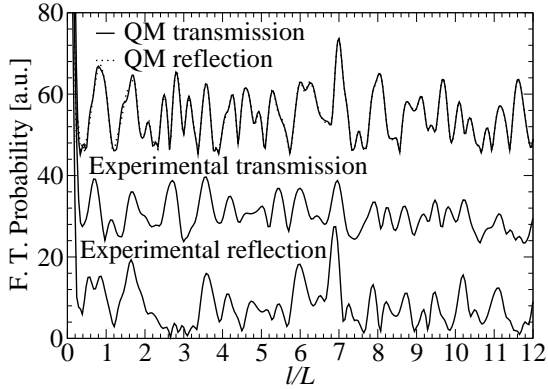}
\caption{\label{fig2} Fourier transform of the
experimental and calculated quantum mechanical (QM)
 transmission and  reflection \textit{probabilities} for the square
 billiard with   opposite leads.}
\end{figure}
%----------------------------------------------------------

Experimental and calculated QM
  transmission and reflection \emph{probabilities} are shown in
  Fig. \ref{fig2}.
The correspondence between the
  theoretical and experimental probabilities is also rather good.
Note that because of the current conservation requirement, $R+T=1$,
  the variation
  of the transmission is opposite to that one of the reflection,
  $\delta T=-\delta R$. As a result, the FT of the \emph{calculated QM}
  transmission and
  reflection probabilities are practically
  identical. This is however not the case for the
  \emph{experimental}  transmission and
  reflection probabilities. This is because of
  the presence of some absorption in
  the system. As we do not include neither absorption nor inelastic
  scattering in  the theoretical calculations, 
 this is the reason for some discrepancy existing
  between  the QM calculations and the experiment.

In contrast to the case of SC and QM \emph{amplitudes}, the
agreement between the SC and QM \emph{probabilities} is only
marginal (therefore we do not show the SC results here).
Because the probabilities are the squared moduli of the
amplitudes, the discrepancy which exists between the SC and QM
\textit{amplitudes}, see Fig. \ref{fig1}, becomes much more
pronounced for the \textit{probabilities} (a detailed analysis of
the discrepancy between the SC and QM approaches is given in
\cite{TB2}). Furthermore,
  the interval of the frequency variation (limited to one propagating
  mode in the leads) is not wide enough to ensure reliable
  statistics for the \textit{probabilities}. The calculations
  demonstrate that with a wider
  frequency interval the characteristic peaks in the FT spectrum
  become better resolved and the agreement between the QM and SC
  results improves significantly. Experimentally, however, it is not
  possible to access the frequency range beyond  one propagating
  mode in the leads.

%-------------------------- Fig 3--------------------------
\begin{figure}
\includegraphics[width=0.4\textwidth]{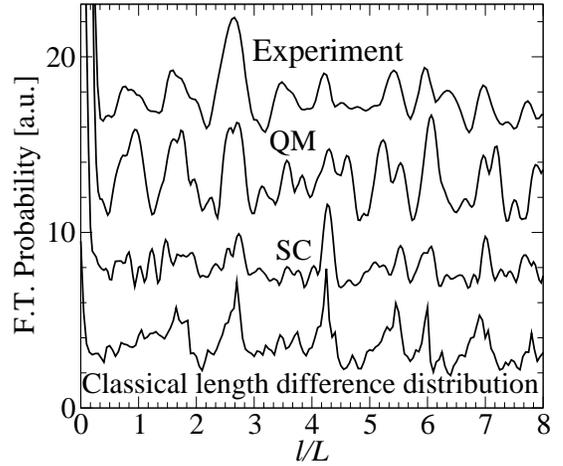}
\caption{\label{fig3} Fourier transform of the
experimental, quantum mechanical (QM) and semiclassical (SC)
transmission {\em probabilities} in a square billiard averaged
over four different lead positions. The characteristic frequencies
in the transmission {\em probabilities} can be understood in terms
of the {\em length difference distribution} in the pairs of
classical trajectories between the leads (the lower
curve). }
\end{figure}
%----------------------------------------------------------

In order to provide the SC interpretation of the probabilities in the
available frequency interval (limited to one propagating mode), we
perform the averaging over four different
lead geometries, see Fig. \ref{fig3}. Such the averaging is justified
because the
characteristic frequencies of the oscillations in a square billiard
are shown to be rather insensitive to the lead positions \cite{TB1}. This, in
turn, is related to the fact that classical length difference distribution
is also not sensitive to the lead positions.
 The averaged QM
  probabilities show pronounced peaks in the FT which are in a good
  agreement with the corresponding experimental ones.
The correspondence between the averaged QM and SC results is also rather
  good. According to the SC approach, the characteristic peaks in the
  SC spectra can be understood in terms of the
length differences in pairs of classical trajectories connecting the
leads, see Eq. (\ref{probab}). This is demonstrated in Fig. \ref{fig3} where the experimental
and calculated spectra are compared to the classical length
  difference distribution between
  the leads.
This provides us with the semiclassical
interpretation of the calculated QM (and therefore observed)
conductance fluctuations. We would like to stress that this
explanation of the characteristic frequencies in the conductance is
based on transport calculations and is thus not equivalent to a
rather common point of view when the observed frequencies in the
conductance oscillations of an open dot are assigned to the
contributions from specific periodic orbits in a corresponding
closed dot
\cite{Marcus,Chang,Persson,Bird,Ensslin,comment,Christ,Bogg,andy_sq,
Takagaki}.

To conclude, we present experimental studies of the geometry-specific
quantum scattering in a microwave billiard of a given shape.  We
perform full quantum
mechanical (QM) scattering calculations and find an excellent agreement
with the experimental results. We also carry out the semiclassical (SC)
calculations where the conductance is given as a sum of all classical
trajectories between the leads, each of them carrying the
quantum-mechanical phase. Our results thus provide the first
{\em unambiguous} identification of the specific frequencies of the
oscillations observed in a billiard of a given shape.

\begin{acknowledgments} Financial support from the National
Graduate School in Scientific Computing (T.B) and the Swedish
Research Council (I.V.Z) is acknowledged. The experiments
were supported by the Deutsche Forschungsgemeinschaft.
\end{acknowledgments}

\end{document}